\documentclass[twocolumn]{aastex63}
\usepackage{xcolor}

\begin{document}

\title{A Model for the Fast Blue Optical Transient AT2018Cow: Circumstellar Interaction of a Pulsational Pair-instability Supernova}

\shorttitle{FBOT AT2018cow: CSM of PPISN}
\shortauthors{Leung, Blinnikov, Nomoto et al.}


\author[0000-0002-4972-3803]{Shing-Chi Leung\thanks{Email address: scleung@caltech.edu}}

\affiliation{Kavli Institute for the Physics and 
Mathematics of the Universe (WPI), The University 
of Tokyo Institutes for Advanced Study, The 
University of Tokyo, Kashiwa, Chiba 277-8583, Japan}

\affiliation{TAPIR, Walter Burke Institute for Theoretical Physics, 
Mailcode 350-17, Caltech, Pasadena, CA 91125, USA}

\author[0000-0002-5726-538X]{Sergei Blinnikov\thanks{Email address: sblinnikov@gmail.com}}

\affiliation{Kavli Institute for the Physics and
Mathematics of the Universe (WPI), The University
of Tokyo Institutes for Advanced Study, The
University of Tokyo, Kashiwa, Chiba 277-8583, Japan}

\affiliation{National Research Center “Kurchatov institute,” Institute for Theoretical and Experimental Physics (ITEP), 117218 Moscow, Russia}

\affiliation{Dukhov Automatics Research Institute (VNIIA), Suschevskaya 22, 127055 Moscow, Russia}

\author[0000-0001-9553-0685]{Ken'ichi Nomoto\thanks{Email address: nomoto@astron.s.u-tokyo.ac.jp}}

\affiliation{Kavli Institute for the Physics and 
Mathematics of the Universe (WPI), The University 
of Tokyo Institutes for Advanced Study, The 
University of Tokyo, Kashiwa, Chiba 277-8583, Japan}

\author[0000-0002-5920-1478]{Petr Baklanov\thanks{Email address: baklanovp@gmail.com}}

\affiliation{National Research Center “Kurchatov institute,” Institute for Theoretical and Experimental Physics (ITEP), 117218 Moscow, Russia}

\affiliation{National Research Nuclear University MEPhI, 
Kashirskoe sh. 31, Moscow 115409, Russia}

\affiliation{Space Research Institute (IKI), Russian Academy of Sciences, Profsoyuznaya 84/32, 117997 Moscow, Russia}

\author{Elena Sorokina\thanks{Email address:sorokina@sai.msu.su}}

\affiliation{National Research Center “Kurchatov institute,” Institute for Theoretical and Experimental Physics (ITEP), 117218 Moscow, Russia}

\affiliation{Sternberg Astronomical Institute, M.V. Lomonosov Moscow
State University, Universitetski pr. 13, 119234 Moscow, Russia}

\author[0000-0002-4587-7741]{Alexey Tolstov\thanks{Email address:alexey.tolstov@ipmu.jp}}

\affiliation{Kavli Institute for the Physics and 
Mathematics of the Universe (WPI), The University 
of Tokyo Institutes for Advanced Study, The 
University of Tokyo, Kashiwa, Chiba 277-8583, Japan}

\affiliation{The Open University of Japan, 2-11, Wakaba, Mihama-ku, Chiba, Chiba
261-8586, Japan}

\correspondingauthor{Shing-Chi Leung}
\email{scleung@caltech.edu}

\received{7 February 2020}
\accepted{18 September 2020}
\published{3 November 2020}
\date{\today}

\begin{abstract}

The Fast Blue Optical Transient (FBOT)
ATLAS18qqn (AT2018cow) has a light curve as bright as superluminous supernovae
but rises and falls much faster. 
We model this light curve by circumstellar interaction of a
pulsational pair-instability (PPI) supernova (SN) model
based on our PPISN models studied in previous work.
We focus on the 42 $M_\odot$ He star (core of a 80 $M_{\odot}$ star)
which has circumstellar matter of mass 0.50 $M_\odot$.
With the parameterized mass cut and the kinetic energy of explosion $E$, we perform
hydrodynamical calculations of nucleosynthesis and optical light
curves of PPISN models.  The optical light curve of the
first $\sim$ 20 days of AT2018cow is well-reproduced by the shock
heating of circumstellar matter for the $42 ~M_{\odot}$ He star with
$E = 5 \times 10^{51}$ erg.  After day 20, the light curve is
reproduced by the radioactive decay of 0.6 $M_\odot$ $^{56}$Co, which
is a decay product of $^{56}$Ni in the explosion.  We also examine how
the light curve shape depends on the various model parameters, 
such as CSM structure and composition. 
We also discuss (1) other possible energy sources and their constraints, 
(2) origin of observed high-energy radiation, and (3) how our result
depends on the radiative transfer codes.  
Based on our successful
model for AT2018cow and the model for SLSN with the CSM mass as large
as $20 ~M_\odot)$, we propose the working hypothesis that PPISN
produces SLSNe if CSM is massive enough and FBOTs if CSM is less than
$\sim 1 ~M_\odot$.

\end{abstract}

\keywords{Supernovae(1668) -- Supernova dynamics(1664) -- Concept: Radiative transfer -- Concept: Light curves -- Stellar pulsations(1625)}

\pacs{
26.30.-k,    
}

\section{Introduction}

The Fast Blue Optical Transient (FBOT) 
ATLAS18qqn/AT2018cow (COW) \citep{Prentice2018,Perley2018}
has a peak luminosity of $1.7 \times 10^{44}$ erg s$^{-1}$, 
being as high as superluminous supernovae (SLSNe) 
\cite[e.g.,][]{GalYam2012}.
But it shows a much faster evolution in its optical properties than
SLSNe.  Its brightness rises five magnitudes within the first three
days and also falls much faster than SLSNe.  It shows hot blackbody
spectra with an effective temperature $\sim 27000$ K, which drops by
$\sim 12000$ K in two weeks.  Its spectra are featureless without
metal lines in both optical and UV, but
show a quasi-static He feature \citep{Kuin2019,Bietenholz2020}.  The
detection of early X-ray and $\gamma$-ray indicates the possibility of
having an inner energy source \citep[e.g.,][]{Margutti2019}.

Several models for the optical light curve of AT2018cow have
been suggested, including the magnetar \citep[e.g.,][]{Fang2019},
electron capture supernova from an accretion-induced collapse of an
ONeMg white dwarf\footnote{See
\cite{Nomoto1984,Nomoto1991,Zha2019,Leung2020ECSN} for details of
the electron capture supernova model.}  \citep{Lyutikov2019}, and
circumstellar interaction model \citep[e.g.,][]{Fox2019}.  The
circumstellar interaction model has been applied to one of FBOT, KSN
2015K, and successfully reproduces the short timescale of rise and
decline \citep{Rest2018,Tolstov2019}.  In particular,
\cite{Tolstov2019} adopted the electron capture supernova model of a
super-AGB star, which has an optically thick CSM with the mass of as
small as $\sim 0.1 ~M_\odot$.

As an alternative to the supernova model, a tidal disruption event
(TDE) has been proposed as a model for AT2018cow.  See
\cite{Perley2018, Liu2018, Kuin2019} for details.
TDE is capable of reproducing the general $t^{-5/3}$ dependence
in supernova light curve \citep{Goicovic2019}, but it depends
on the exact orbits around the massive black hole. For close
encounter, the strong tidal force can trigger spontaneous 
nuclear runaway and explosion \citep{Tanikawa2018a,Tanikawa2018b}.

In the present work, we consider the circumstellar interaction of the
pulsational pair-instability (PPI) supernova (SN).  It has been shown
that stars as massive as initially 80-140 $M_\odot$ undergo PPI during
O-burning because of the electron-positron pair creation
\citep[e.g.,][]
{Barkat1967,Heger2002,
Ohkubo2009,Yoshida2014,Woosley2017,Marchant2019,Leung2018PPISN1}
 \cite{Woosley2007}
calculated the sequence of pulsational mass ejection of H-rich
materials and interaction between the ejecta during the pulsation.
Then they tried to reproduce the light curve of a Type II SLSN
(SLSN-II) 2006.

\begin{figure}
\centering
\includegraphics*[width=8cm,height=6cm]{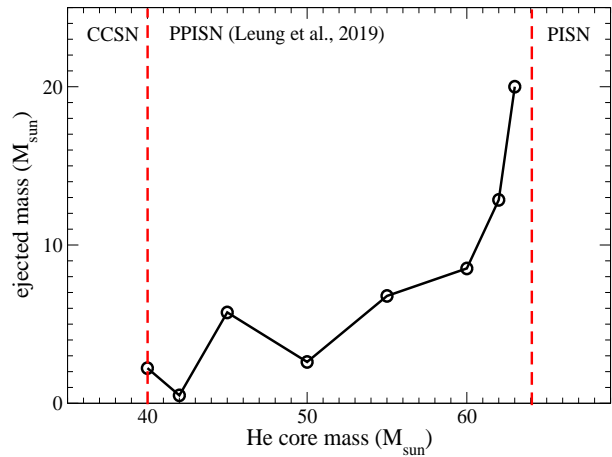}
\caption{Total ejecta mass by pulsation
against progenitor He core mass.}
\label{fig:ejecta_mass_plot}
\end{figure}

In \cite{Leung2018PPISN1}, 
we have calculated the evolution of 80 -- 120 $M_\odot$
stars with the metallicities of $Z = 10^{-3} - 1.0$ $Z_\odot$ through
the beginning of PPI.  Assuming H-rich envelope is lost, we have
further evolved their He cores of 40 -- 62 $M_\odot$ with $Z = 0$ through
the core-collapse.  During the pulsation, these He stars undergo
extensive mass loss.  The ejected masses are 3 -- 13 $M_\odot$ for $40
- 62 ~M_\odot$ He stars, as seen in Figure \ref{fig:ejecta_mass_plot}, because the pulsations
are stronger for more massive He stars \cite[see also][]{
Yoshida2014,Woosley2017,Marchant2019}.  The ejecta form He-rich CSM.

The lack of metal lines in the spectra of AT2018cow suggests that the ejecta
contains mostly He \citep{Prentice2018}.  This feature is consistent
with the PPISN model which ejects mostly the outer layer of the He
star during pulsation.  (The exact composition can be affected by
other stellar parameters such as the progenitor mass and rotation
\citep{Chatzopoulos2012}.)

\citet{Tolstov2017} applied a PPISN model of 50 $M_\odot$ He star
(which is a He core of 100 $M_\odot$ star) with a large amount of an
optically thick CSM $(\sim 20 ~M_\odot)$ and the kinetic energy of
explosion as high as $\sim 10^{52}$ erg s$^{-1}$.  They calculated the
circumstellar interaction and the resultant light curve.  The model
well-explains the light curve of Type I SLSN (SLSN-I) PTF 12dam, whose
early curve shows a rise of 2.5 mag in 20 days.  Such massive CSM is
good to reproduce the slow rise of SLSN-I (see also
\citep{Sorokina2016}), but too massive to explain the fast rise of the
light curve of AT2018cow (5 mag in 5 days).

\cite{Perley2018} have suggested that in order to reproduce
the light curve qualitatively of AT2018cow, 
there is a pre-explosion ejected mass of $\sim$0.5 $M_{\odot}$.
We thus look for in Figure \ref{fig:ejecta_mass_plot} the pulsational 
pair-instability supernova model
which produces a similar CSM mass. We find that this
corresponds to the He star model with the mass 
$M_{{\rm He}} = 42 ~M_{\odot}$. 
This He star ejects $0.50 ~M_{\odot}$ of its surface matter, composed of
only He, to the surrounding at $\sim 1.6$ year prior to its final
collapse.

As seen from the models for SLSN \citep{Tolstov2017} and
FBOT \citep{Tolstov2019}, the mass of the optically thick CSM seems to
determine the rising timescale of light curves.  Based on this, we set
here the following working hypothesis that PPISN produces SLSNe if CSM
is massive enough and FBOTs if CSM is less than $\sim 1 M_\odot$.

Based on our working hypothesis, we perform the hydrodynamical
simulations of the PPISN explosion of the $42 ~M_{\odot}$ He star.  We
assume that the He star undergoes core collapse to form a black hole
and generates an explosion.  With the parameterized explosion energy, we
calculate nucleosynthesis, circumstellar interaction, and bolometric
light curves.  From the comparison with the observed light curve of
AT2018cow, we obtain the constraints on the explosion energy, mixing
of radioactive $^{56}Ni$, density distribution of CSM, and some other
model parameters.
  
We note that radio and X-ray observations \citep{Margutti2019}
have provided hints that aspherical explosion is necessary
for the peculiar evolution of this object. In this work, we use the 
one-dimensional simulation with spherical approximation 
as an exploratory work to see which parameters are necessary for 
reproducing the light curve of this object. We aim for fitting 
the global features of this light curve.

In Sections \ref{sec:methods} and \ref{sec:nucleo}
we review the numerical scheme to
calculate nucleosynthesis and the bolometric light curve.

In Section \ref{sec:csmInteraction} we describe our optimized model which has a
bolometric light curve closet to AT2018cow.  We present the
detailed evolution of the hydrodynamics and radiative transfer after
its explosion.

In Section \ref{sec:sensitivity} we present a detailed numerical study
to examine the sensitivity of our results on the model parameter and
input physics. This includes the explosion energy, $^{56}$Ni
distribution, and CSM properties.

In Section \ref{sec:discussion} we further discuss how other central
energy sources can contribute or supplement to the light curve of
AT2018cow.  We also discuss how such central energy sources contribute
produce to the high-energy photons and compare with AT2018cow.  In the
end, we show how our results depend on the radiative transfer code.

\section{Initial Model and Methods of Hydrodynamical Simulations}
 \label{sec:methods}
  
\subsection{Presupernova Model}

As discussed in Introduction, we adopt the $42 ~M_{\odot}$ He star
model which was evolved from the He main-sequence to the Fe core
collapse by \cite{Leung2018PPISN1} through PPI and associated mass
ejection. We used the code MESA (Modules for the Experiments in Stellar
Astrophysics) \citep{Paxton2011,Paxton2013,Paxton2015,Paxton2017}
(Version 8118).

\begin{figure*}
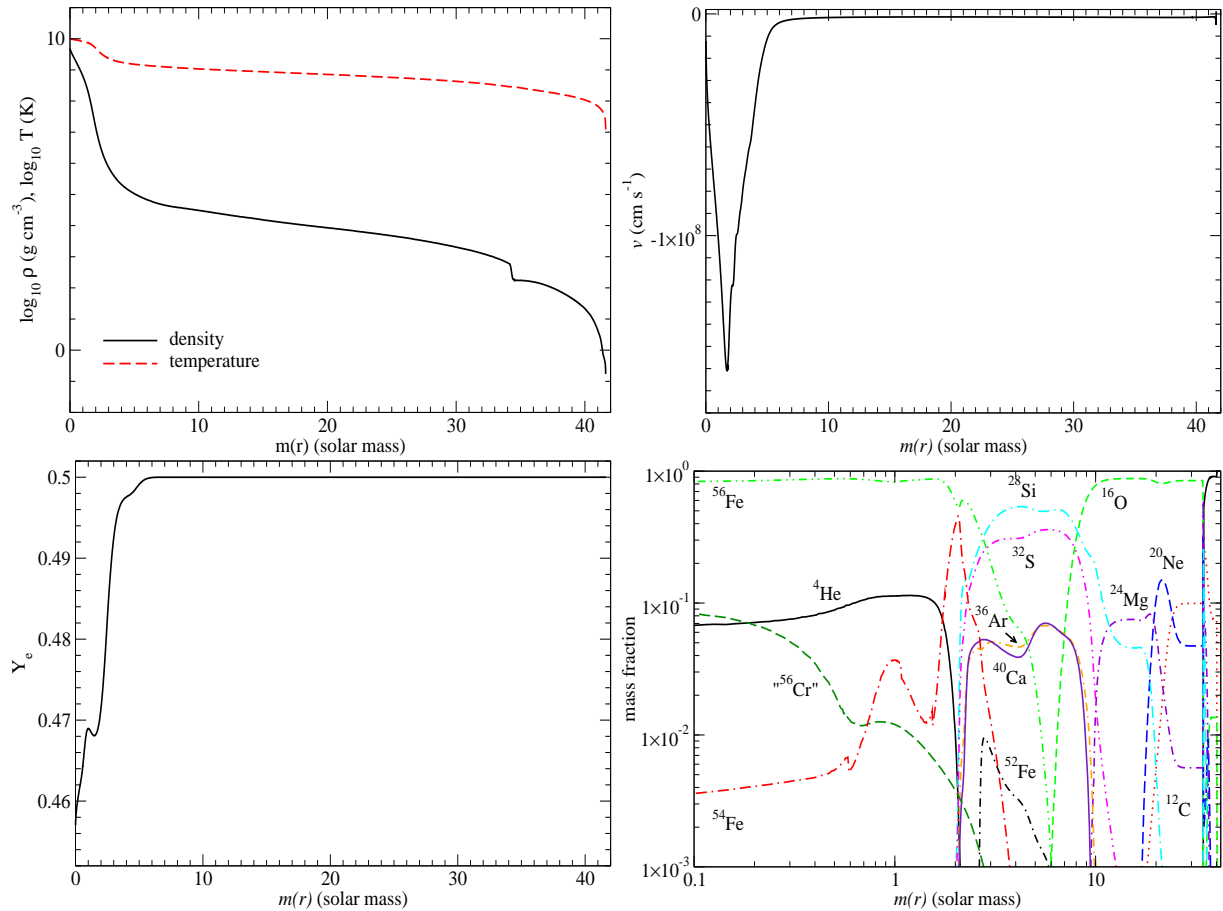

\centering
\includegraphics*[width=8cm,height=6cm]{fig1a.eps}
\includegraphics*[width=8cm,height=6cm]{fig1b.eps}
\includegraphics*[width=8cm,height=6cm]{fig1c.eps}
\includegraphics*[width=8cm,height=6cm]{fig1d.eps}
\caption{The density and temperature (top left), 
velocity (top right), electron mole fraction (bottom left) and
abundance (bottom right) of the supernova model
at the onset of collapse.}
\label{fig:He42_plot}
\end{figure*}

In Figure \ref{fig:He42_plot} we plot the pre-explosion configurations
for the density and temperature (top left panel),
velocity (top right panel), electron mole fraction $Y_{\rm e}$ (bottom left
panel), and the chemical abundance (bottom right panel) 
against $M(r)$, where $M(r)$ is the included mass within the radius $r$.
Because the He star has lost 0.50 $M_{\odot}$ during PPI, 
the mass of the collapsing He star is 41.50 $M_{\odot}$.

The chemical abundance profile shows three layers.
The inner core at $M(r) < 2 M_\odot$, which collapses to a proto-neutron star (and later a black hole), is made of mostly $^{56}$Fe. 
A small fraction of $^{4}$He exists due to the
photo-disintegration. A trace of electron capture 
products, represented by $^{56}$Cr
\footnote{In the default setting of MESA \citep{Paxton2017},
the isotope $^{56}$Cr ($Y_{\rm e} = 0.43$)
is used as a representative of electron capture
products when the stellar core reaches a temperature
sufficient for nuclear statistical equilibrium.
This can keep the stellar evolutionary model achievable in 
a reasonable computational time.  
We remind that in general a wide range of isotopes
with different neutron-proton ratios is necessary to capture the 
self-consistent electron capture rate. However, a large
reaction network can make the general hydrodynamical 
treatment numerically demanding.}, exist in the core, 
being insignificant at $M(r) > 1 ~M_{\odot}$.
The envelope includes an inner envelope and an outer
envelope which differ by their compositions. 
The inner envelope at $M(r) = 2 - 8 ~M_{\odot}$
consists of mostly intermediate mass elements, featured
by $^{28}$Si and $^{32}$S, and smaller amounts of 
$^{36}$Ar and $^{40}$Ca. In the outer envelope 
at $M(r) = 8 - 35 ~M_{\odot}$, 
$^{16}$O dominates the composition, with a trace amount of
$^{20}$Ne and $^{24}$Mg.
The transition to the outer shell around $M(r) \sim 35 M_{\odot}$ can be observed by the emergence of
$^{28}$Si near the former shock breakout region during pulsation. 
The outer shell at $M(r) > 35 ~M_{\odot}$ is mostly
$^{4}$He with a trace amount of $^{12}$C.

The $Y_{\rm e}$ profile shows $Y_{\rm e} \sim 0.46$ in the Fe core 
and sharp transition to $Y_{\rm e} = 0.5$ at $M(r) > 5~M_{\odot}$.

The density profile reveals a three-layer structure of the 
pre-supernova: the Fe-core at $M(r) <  2~ M_{\odot}$, the Si-O envelope
at $M(r) = 2 - 35 ~M_{\odot}$, and the outer He-shell at $M(r) > 35 ~M_{\odot}$.  
On the other hand, the temperature profile shows only the core and the envelope
structure with no sharp gradient around the He-O interface.
The Fe-core features a sharp drop of density 
of 5 orders of magnitude. 
The Si-O envelope has a shallow density gradient 
from $10^5$ to $10^3$ g cm$^{-3}$.
The transition
to the outer He-shell is shown by the sharp density change.
The density also drops sharply towards the surface.

The velocity profile in the Fe core shows homologous contraction 
at $M(r) < 2 M_{\odot}$ with a peak velocity
at $1.8 \times 10^8$ cm s$^{-1}$. Then the velocity
is zero at $M(r) \sim 5 M_{\odot}$. At $M(r) > 5 ~M_{\odot}$
the star is close to hydrostatic.

\subsection{Methods of Nucleosynthesis and Radiative Transfer}

For the progenitor model as described in \ref{sec:methods}, 
we first perform the
hydrodynamical calculations of nucleosynthesis using the
one-dimensional code described in \cite{Tominaga2007}.
The shock wave is calculated with the PPM method. 
Nucleosynthesis is calculated with the $\alpha$-network (13 species) coupled with the
hydrodynamics and with a big network (280 species) for a post-processing.

For the bolometric light curve calculation, we use the radiation hydrodynamics
code (SNEC) \citep{Morozova2015}. 
The code calculates the transport of blackbody radiation in the diffusion limit 
to obtain the bolometric light curve. The opacity takes the
Rosseland mean opacity mainly from the OPAL
tables \citep{Iglesias1993,Iglesias1996}.  The ionization is obtained
by solving the Saha equation. For the equation of state (EOS), the
Paczynski EOS \citep{Paczynski1983} is adopted.
The gamma-ray heating from the decays of radioactive $^{56}$Ni and
$^{56}$Co is calculated assuming the gray transfer approximation and
pure absorptive opacity.

\section{Explosive Nucleosynthesis and Radioactive Heating}
\subsection{Explosive Nucleosynthesis}
\label{sec:nucleo}

For the optimized model in this paper, 
we set a thermal bomb at the mass cut at $M_{\rm cut} = 2.0~
M_{\odot}$ with the internal energy to make a kinetic energy of
explosion (hereafter "explosion energy") $E = 5 \times 10^{51}$ erg s$^{-1}$.
We then calculate the propagation of a shock wave coupled with
nucleosynthesis.
Nuclear energy release is added to $E$ but negligibly small.

\begin{figure}
\centering
\includegraphics*[width=8cm,height=6cm]{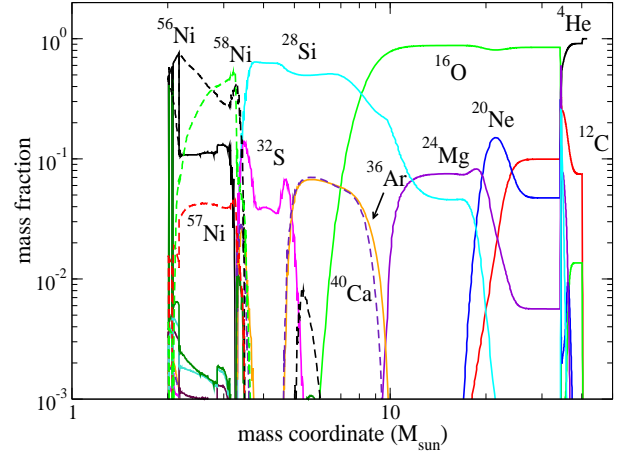}
\caption{The distributions of the chemical abundances 
  at $t =$ 50 s after the thermal bomb is deposited at
  $M_{\rm cut} = 2.0 M_{\odot}$ to produce $E = 5 \times 10^{51}$ erg s$^{-1}$.
}
\label{fig:He42_postexp_plot}
\end{figure}

Figure \ref{fig:He42_postexp_plot} shows the distributions of the
post-shock temperature and density at $t =$ 50 s, 
where $t$ denotes the time after the thermal bomb is deposited.
$Y_{\rm e}$ does not change from the presupernova values because of
the low density in the ejecta.

The core from $M_{\rm cut}$ to $\sim 3~M_{\odot}$ contains iron-peak
nuclei including mostly $^{56-58}$Ni and $^{4}$He synthesized in the
$\alpha$-rich freezeout.  The middle
layer at $M(r) = \sim 3 - 10~M_{\odot}$ is occupied by intermediate mass species including $^{28}$Si,
$^{32}$S, $^{36}$Ar and $^{40}$Ca.  The inner envelope at $M(r) \sim 10 - 35~ M_{\odot}$ 
contains mostly light elements mostly $^{16}$O,
and then $^{20}$Ne and $^{24}$Mg.  At the surface there are unburned
$^{4}$He and a small fraction of $^{12}$C.
In Table \ref{table:abundance}, we summarize nucleosynthesis yields of radioactive nuclei at $t =$ 50 s
and several species after radioactive decays.
In particular, the amount of $^{56}$Ni is 0.62 $M_{\odot}$.

\begin{table}
\begin{center}
\label{table:abundance}
\caption{The ejecta mass and scaled abundance fraction of the star after explosive nucloesynthesis where all exothermic nuclear reaction has stopped. All masses are in units of $M_{\odot}$.}
\begin{tabular}{|c|c|c|c|c|}
\hline
Isotope & $A$ & $Z$ & $(X_i/^{56}$Fe)/$(X_i/^{56}$Fe)$_{\odot}$ & Mass \\ \hline
  $^{12}$C & 12 & 6 & $1.06$ & 1.91 \\
  $^{13}$C & 13 & 6 & $2.11 \times 10^{-13}$ & $4.55 \times 10^{-15}$ \\ \hline

  $^{14}$N & 14 & 7 & $1.10 \times 10^{-8}$ & $4.77 \times 10^{-9}$ \\
  $^{15}$N & 15 & 7 & $5.07 \times 10^{-10}$ & $8.72 \times 10^{-13}$ \\ \hline

  $^{16}$O & 16 & 8 & $5.56$ & 22.07 \\
  $^{17}$O & 17 & 8 & $7.76 \times 10^{-11}$ & $1.23 \times 10^{-13}$ \\
  $^{18}$O & 18 & 8 & $1.12 \times 10^{-8}$ & $1.02 \times 10^{-10}$ \\ \hline
  
  $^{19}$F & 19 & 9 & $2.11 \times 10^{-9}$ & $4.21 \times 10^{-13}$ \\ \hline

  $^{20}$Ne & 20 & 10 & $1.88$ & 1.174 \\
  $^{21}$Ne & 21 & 10 & $1.09 \times 10^{-9}$ & $2.17 \times 10^{-12}$ \\
  $^{22}$Ne & 22 & 10 & $1.24 \times 10^{-10}$ & $1.04 \times 10^{-11}$ \\ \hline

  $^{23}$Na & 23 & 11 & $1.81 \times 10^{-7}$ & $3.36 \times 10^{-9}$ \\ \hline

  $^{24}$Mg & 24 & 12 & $3.84$ & 1.04 \\
  $^{25}$Mg & 25 & 12 & $7.93 \times 10^{-7}$ & $2.86 \times 10^{-8}$ \\
  $^{26}$Mg & 26 & 12 & $2.34 \times 10^{-5}$ & $9.68 \times 10^{-7}$ \\ \hline

  $^{27}$Al & 27 & 13 & $3.43 \times 10^{-4}$ & $1.06 \times 10^{-5}$ \\ \hline

  $^{28}$Si & 28 & 14 & $ 10.12$ & 3.52 \\
  $^{29}$Si & 29 & 14 & 0.517 & $9.51 \times 10^{-3}$ \\
  $^{30}$Si & 30 & 14  & $1.11 \times 10^{2}$ & 1.39 \\ \hline

  $^{31}$P & 31 & 15 & $2.96 \times 10^{-4}$ & $8.05 \times 10^{-7}$ \\ \hline

  $^{32}$S & 32 & 16 & 0.515 & 0.106 \\
  $^{33}$S & 33 & 16 & 2.81 & $4.75 \times 10^{-3}$ \\
  $^{34}$S & 34 & 16 & 18.00 & 0.174 \\
  $^{36}$S & 36 & 16 & 0.696 & $2.30 \times 10^{-5}$ \\ \hline

  $^{35}$Cl & 35 & 17 & 0.218 & $3.68 \times 10^{-4}$ \\
  $^{37}$Cl & 37 & 17 & 0.489 & $2.83 \times 10^{-4}$ \\ \hline
  
  $^{36}$Ar & 36 & 18 & 5.17 & 0.224 \\
  $^{38}$Ar & 38 & 18 & 4.73 & $4.05 \times 10^{-2}$ \\
  $^{40}$Ar & 40 & 18 & $2.30 \times 10^{-2}$ & $2.48 \times 10^{-6}$ \\ \hline

  $^{39}$K & 39 & 19 & 0.151 & $2.60 \times 10^{-4}$ \\
  $^{40}$K & 40 & 19 & 0.213 & $5.54 \times 10^{-7}$ \\
  $^{41}$K & 41 & 19 & 1.50 & $2.00 \times 10^{-4}$ \\ \hline

  $^{40}$Ca & 40 & 20 & 6.55 & 0.214 \\
  $^{42}$Ca & 42 & 20 & 1.06 & $2.40 \times 10^{-4}$ \\
  $^{43}$Ca & 43 & 20 & 0.257 & $1.36 \times 10^{-5}$ \\
  $^{44}$Ca & 44 & 20 & 2.53 & $1.94 \times 10^{-3}$ \\
  $^{46}$Ca & 46 & 20 & $2.91 \times 10^{-3}$ & $3.72 \times 10^{-9}$ \\
  $^{48}$Ca & 48 & 20 & $9.04 \times 10^{-10}$ & $6.80 \times 10^{-14}$ \\ \hline

  $^{45}$Sc & 45 & 21 & $8.94 \times 10^{-2}$ & $1.68 \times 10^{-6}$ \\ \hline

  $^{46}$Ti & 46 & 22 & 0.462 & $5.45 \times 10^{-5}$ \\
  $^{47}$Ti & 47 & 22 & 0.554 & $6.15 \times 10^{-5}$ \\
  $^{48}$Ti & 48 & 22 & 1.99 & $2.29 \times 10^{-3}$ \\
  $^{49}$Ti & 49 & 22 & 0.418 & $3.65 \times 10^{-5}$ \\
  $^{50}$Ti & 50 & 22 & 0.223 & $1.93 \times 10^{-5}$ \\ \hline

  $^{50}$V & 50 & 23 & 7.60 & $3.13 \times 10^{-6}$ \\
  $^{51}$V & 51 & 23 & 3.86 & $6.71 \times 10^{-4}$ \\ \hline

\end{tabular}
\end{center}
\end{table}

\begin{table}
\begin{center}
\label{table:abundance2}
\caption{\textit{(cont'd)} The ejecta mass and scaled abundance fraction of the star after explosive nucloesynthesis where all exothermic nuclear reaction has stopped. All masses are in units of $M_{\odot}$. $A$ and $Z$ are the atomic mass and number of the isotopes.}
\begin{tabular}{|c|c|c|c|c|}
\hline
Isotope & $A$ & $Z$ & $(X_i/^{56}$Fe)/$(X_i/^{56}$Fe)$_{\odot}$ & Mass \\ \hline

  $^{50}$Cr & 50 & 24 & 9.26 & $3.41 \times 10^{-3}$ \\
  $^{52}$Cr & 52 & 24 & 7.38 & $5.49 \times 10^{-2}$ \\
  $^{53}$Cr & 53 & 24 & 2.50 & $2.16 \times 10^{-3}$ \\
  $^{54}$Cr & 54 & 24 & 0.492 & $1.08 \times 10^{-4}$ \\ \hline

  $^{55}$Mn & 55 & 25 & 1.56 & $1.07 \times 10^{-2}$ \\ \hline

  $^{54}$Fe & 54 & 26 & 4.73 & 0.181 \\
  $^{56}$Fe & 56 & 26 & 1.00 & 0.623 \\
  $^{57}$Fe & 57 & 26 & 3.26 & $4.94 \times 10^{-2}$ \\
  $^{58}$Fe & 58 & 26 & $9.29 \times 10^{-3}$ & $2.16 \times 10^{-5}$ \\ \hline

  $^{59}$Co & 59 & 27 & 1.41 & $2.47 \times 10^{-3}$ \\ \hline

  $^{58}$Ni & 58 & 28 & 14.74 & 0.374 \\
  $^{60}$Ni & 60 & 28 & 2.01 & $2.04 \times 10^{-2}$ \\
  $^{61}$Ni & 61 & 28 & 4.70 & $2.20 \times 10^{-3}$ \\
  $^{62}$Ni & 62 & 28 & 41.51 & $6.07 \times 10^{-2}$ \\
  $^{64}$Ni & 64 & 28 & $1.41 \times 10^{-8}$ & $6.29 \times 10^{-12}$ \\ \hline

  $^{63}$Cu & 63 & 29 & 0.474 & $1.50 \times 10^{-4}$ \\
  $^{65}$Cu & 65 & 29 & $2.28 \times 10^{-4}$ & $3.34 \times 10^{-8}$ \\ \hline

  $^{64}$Zn & 64 & 30 &  $1.66 \times 10^{-2}$ & $8.77 \times 10^{-6}$ \\
  $^{66}$Zn & 66 & 30 &  $1.86 \times 10^{-6}$ & $5.60 \times 10^{-10}$ \\
  $^{67}$Zn & 67 & 30 &  $7.11 \times 10^{-8}$ & $3.29 \times 10^{-12}$ \\
  $^{68}$Zn & 68 & 30 &  $1.11 \times 10^{-8}$ & $2.38 \times 10^{-12}$ \\
  $^{70}$Zn & 70 & 30 & $1.48 \times 10^{-7}$ & $1.09 \times 10^{-12}$ \\ \hline

\end{tabular}
\end{center}
\end{table}

\subsection{Radioactive Decays and Light Curve}

\begin{figure}
\centering
\includegraphics*[width=8cm,height=6cm]{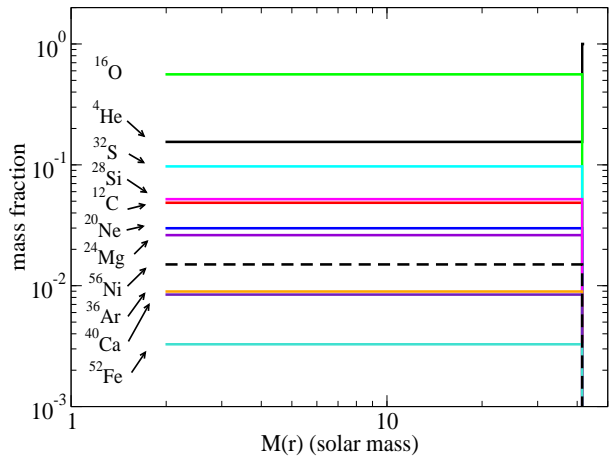}
\caption{The abundance pattern of the ejecta and CSM adopted in the 
optimized model (See next plot). The ejecta is assumed to be 
completely mixed as a representation of aspherical explosion.}
\label{fig:uniform_mix}
\end{figure}

\begin{figure}
\centering
\includegraphics*[width=8cm,height=6cm]{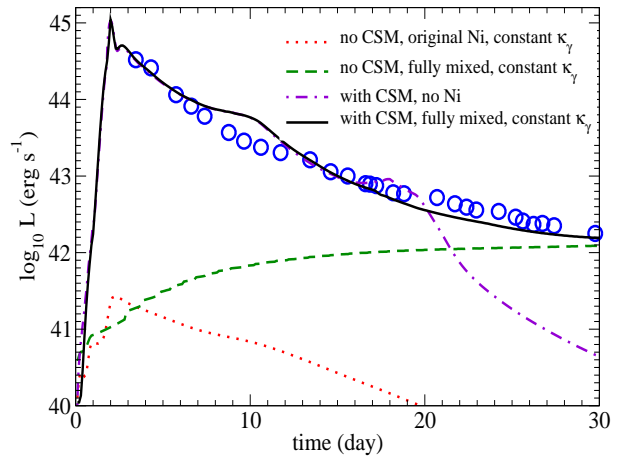}
\caption{Bolometric light curves powered by the radioactive decays of $^{56}$Ni
and $^{56}$Co for various abundance distributions are shown to compare
with the data points (blue circles) observed from AT2018cow
\citep{Perley2018}.  The red dotted line and green dashed line show
the light curve models for the original $^{56}$Ni distribution and the
uniformly mixed one, respectively.  Here no circumstellar interaction
is included.  For gamma-ray transport, $\kappa_{\gamma} = 0.06 Y_{\rm
  e}$ g cm$^{-2}$ is adopted.  For comparison, the black solid
line shows the light curve powered by circumstellar interaction but no
$^{56}$Ni in the ejecta. The purple dot-dash line shows a similar 
light curve but assumes all isotopes in the ejecta is fully mixed,
including $^{56}$Ni.}
\label{fig:benchmark_kappa_plot}
\end{figure}

In the adopted PPISN model, the power sources of the optical light
curve are circumstellar interaction (Section \ref{sec:csmInteraction})
and radioactive decays of $^{56}$Ni and $^{56}$Co.  In Figure
\ref{fig:benchmark_kappa_plot}, several theoretical LCs are compared
with the observed light curve of AT2018cow \citep{Perley2018}.
For the radioactive decay light curve models in Figure
\ref{fig:benchmark_kappa_plot}, we adopt $\kappa_{\gamma} =$ 0.06
$Y_{\rm e}$ cm$^2$ g$^{-1}$.

(1) The dash-dotted purple curve is the bolometric light curve
powered by circumstellar interaction only (see section \ref{sec:csmInteraction})
without radioactive decays.
The opacity are calculated for the original abundance
distribution (Fig. \ref{fig:He42_postexp_plot}).
This light curve can reproduce the observed light curve for only the first 20 days.

(2) The dotted-red curve shows the bolometric light
curve powered by radioactive decays without circumstellar
interaction for the original
(centered) distribution of 0.62 $M_\odot$ $^{56}$Ni (Fig. \ref{fig:He42_postexp_plot}).  
It's peak luminosity reaches only $\sim 10^{41.5}$ erg s$^{-1}$, which is too
faint to explain the observations.  This implies it takes too long
time for radioactive heat to reach the surface because of massive
ejecta.

(3) The above comparison suggests that extensive mixing of $^{56}$Ni to the surface
takes place possibly by a jet-like explosion.  We thus make a simple
assumption that the ejecta is uniformly mixied from the mass cut of $M(r) =
2~M_{\odot}$ to the surface of $M(r) = 41.60~M_{\odot}$.
The uniform abundance distribution is shown in Figure \ref{fig:uniform_mix}.
The calculated bolometric light curve is shown by the dashed-green curve.
Thanks to the $^{56}$Ni heating in the outer layers, the light curve is
consistent with the observations at $t > 25$ days.

(4) The solid-black curve shows the bolometric light curve with the
combined powers of circumstellar interaction and radioactive decays,
The uniform abundance distribution is adopted (Figure
\ref{fig:uniform_mix}) to calculate the radioactive heating and opacity.
The calculated light curve is in good agreement with the
observed light curve of AT2018cow.
The later light curve at $t > $ 20 days declines slower than
AT2018cow, which will be discussed in section.
We thus adopt the uniform abundance distribution in Figure
\ref{fig:uniform_mix} for the model which we call the ``optimized''
model.

\section{Circumstellar Interaction}
\label{sec:csmInteraction}

\subsection{Formation and Structure of Circumstellar Matter}
\label{sec:csmFormation}

In calculating the light curve powered by circumstellar interaction,
we adopt the He star model of $M_{\rm He} = 42.10 ~M_\odot$, which
undergoes PPI and ejects He-rich surface materials to form CSM of mass
$M_{\rm CSM} = 0.50 M_\odot$ at $\sim 1.6$ year prior to its collapse
(see Introduction).  Thus the He star has $M_{\rm He} = 41.60 M_\odot$
at the beginning of Fe core collapse.

\begin{figure}
\centering
\includegraphics*[width=8cm,height=6cm]{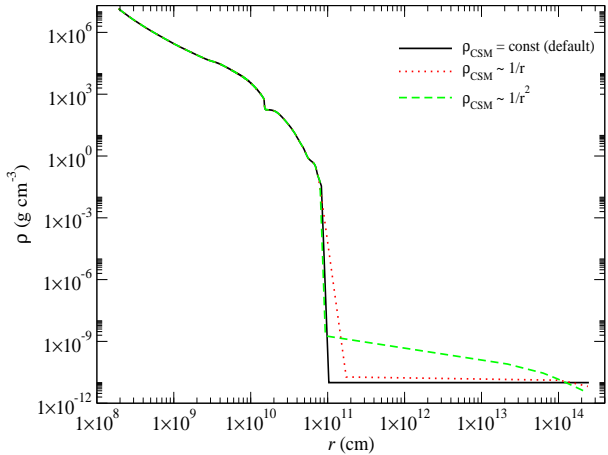}
\caption{The stellar and CSM density profiles of the initial models.
The black-solid line shows the optimized model.  Other lines
  show those for comparisons in section \ref{sec:sensitivity}.
  All models have the same CSM mass $M_{\rm CSM} = 0.50 ~M_{\odot}$ and
  an outer CSM radius $R_{\rm CSM} \sim 10^{14}$ cm.}
\label{fig:rho_ini_tail_plot}
\end{figure}

We plot in Figure \ref{fig:benchmark_kappa_plot} the bolometric light curve using the chemical distribution shown in Figure \ref{fig:uniform_mix}. We also present contrasting models to demonstrate the individual contributions of CSM and $^{56}$Ni-decay. 
We assume that CSM has a constant density of $\rho_{\rm CSM} =
10^{-11}$ g cm$^{-3}$ extending to $R_{\rm CSM} \sim 10^{14}$ cm as
seen in Figure \ref{fig:rho_ini_tail_plot}.  Note such CSM is
optically thick.  We examine in section \ref{sec:sensitivity} how the
shape of the bolometric light curve depends on $M_{\rm CSM}$,
$\rho_{\rm CSM}$, and $R_{\rm CSM}$, and how the comparison with
AT2018cow provides constraints on these quantities.

Spectroscopic observations of AT2018cow have reported the appearance
of H-features in the spectra $\sim$ 15 days after the light maximum
\citep{Perley2018}.  The existence of some H in the ejecta and CSM can
be explained as follows.  During the evolution, the progenitor star of
the PPISN loses a large fraction of its massive H envelope
\citep[e.g.,][]{Leung2018PPISN1}.  The exact amount of H which remains
in the progenitor depends strongly on the mass, metallicity, and
binarity of the progenitor.  If some amount of H remains in the star
when the mass ejection due to PPI occurs, CSM of the PPISN contains
some H.

Another possibility that some H exist in CSM is the case where 
H-rich environment has been formed outside the progenitor star.
During the ejection of the He envelope due to PPI, the high velocity
He-shell may interact with the surrounding low velocity H-rich
materials.  The deceleration of the He shell causes Rayleigh-Taylor
instabilities and some H-rich matter is mixed into the He shell.

We also examine in section \ref{sec:sensitivity} how the existence of
H affacts the light curve shape

\subsection{Hydrodynamical Evolution of Shock Propagation}

We start the radiation hydrodyanamical simulation from the Fe core
collapse.  A shock wave is generated by inserting a thermal bomb at
the mass cut $M_{\rm cut} = 2 ~M_{\odot}$ with the explosion energy $E
= 5 \times 10^{51}$ erg.  The elemental abundance profile obtained
from the explosive nucleosynthesis calculation (Figure
\ref{fig:He42_postexp_plot}) is assumed to be uniformly mixed (Figure
\ref{fig:uniform_mix}).

\begin{figure*}
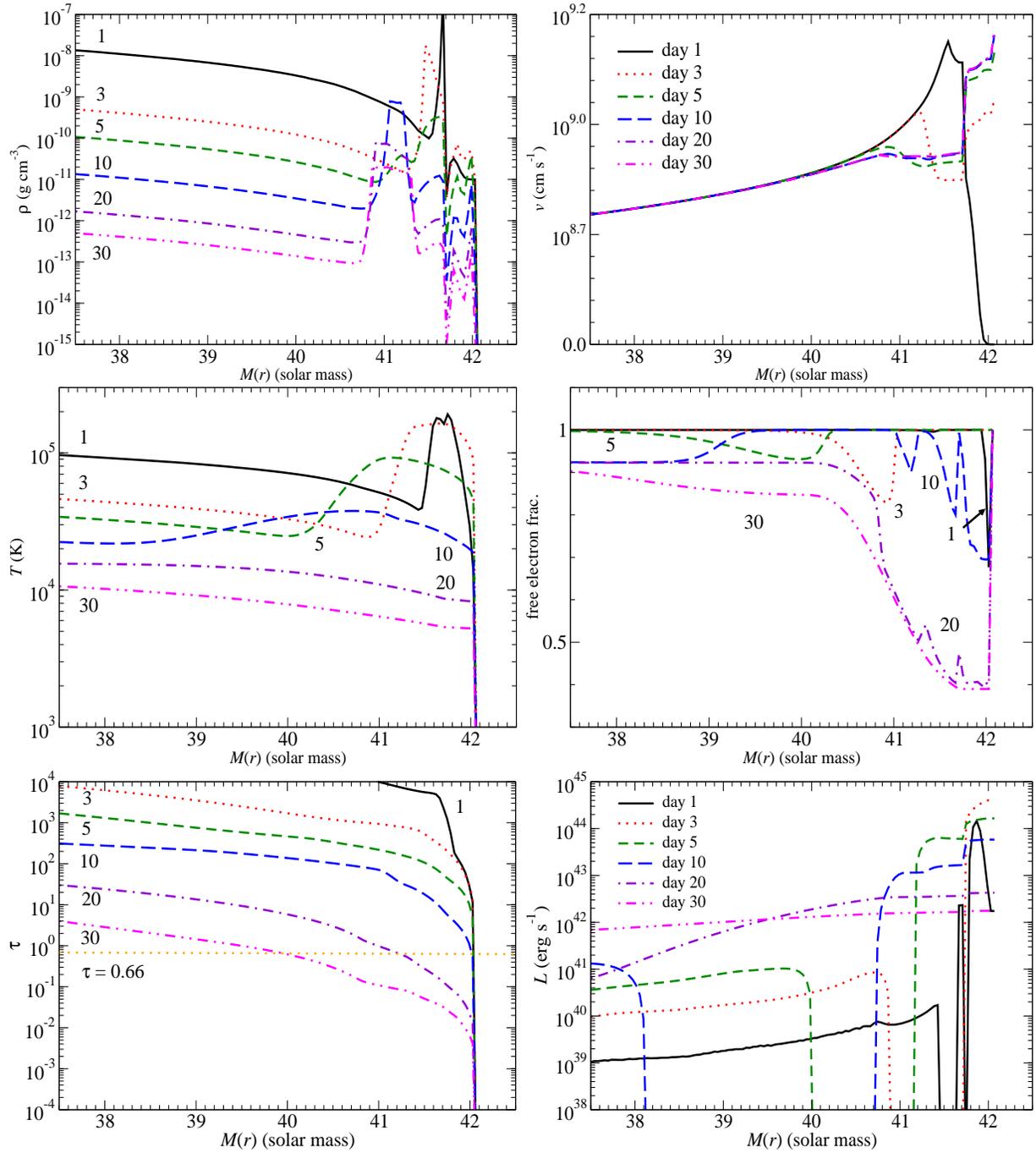

\centering
\includegraphics*[width=8cm,height=6cm]{rho_profiles_plot.eps}
\includegraphics*[width=8cm,height=6cm]{vel_profiles_plot.eps}
\includegraphics*[width=8cm,height=6cm]{temp_profiles_plot.eps}
\includegraphics*[width=8cm,height=6cm]{freee_profiles_plot.eps}
\includegraphics*[width=8cm,height=6cm]{tau_profiles_plot.eps}
\includegraphics*[width=8cm,height=6cm]{lumin_profiles_plot.eps}
\caption{The density (top left), velocity (top left),
temperature (middle left), 
free electron fraction (middle right), optical depth (bottom left) and
luminosity (bottom right) of the benchmark model at
Day 1 (solid black line), 1
(red dotted line), 5 (green dashed line), 10 (blue long-dashed line), 
20 (purple dot-dash line) and 30 (magenta dot-dot-dash line)
after the injection of energy.}
\label{fig:benchmark_hydro_plot}
\end{figure*}

In Figure \ref{fig:benchmark_hydro_plot} we plot the distributions of
the density (top left panel), velocity (top right panel),
temperature (middle left), free electron fraction (middle right),
optical depth (bottom left) and luminosity profile (bottom right)
for the optimized model at Day 0 (solid black line), 1
(red dotted line), 5 (green dashed line), 10 (blue long-dashed line), 
20 (purple dot-dash line) and 30 (cyan dot-long dash line)
after the formation of the shock wave.
Here CSM at $M(r) = 41.6 - 42.1 ~M_\odot$ is included. 

\textit{Density ($\rho$)}: The initial density
profile is the pre-collapse profile shown in Figure \ref{fig:He42_plot}
and the $0.50 ~M_{\odot}$ CSM with a constant density of $10^{-11}$ 
g cm$^{-3}$ extending to $\sim 10^{14}$ cm as seen in Figure \ref{fig:rho_ini_tail_plot}.
We note that CSM is optically thick.
At Day 1, the shock wave has arrived at the inner radius of
CSM and enhanced the density there by two orders of magnitude.
At Day 2, the shock-breakout from CSM occurs. 
At Day 5, the post-shock structure develops with a trough 
inside the inner CSM and a bump just behind CSM. 
A reverse shock can be seen around Day 10, 
but it freezes when the free expansion of matter dominates the motion.

\textit{Velocity (v)}: When the shock wave arrives at CSM at Day 1, 
the velocity at the inner edge of CSM reaches as 
high as $1.4 \times 10^{9}$ cm s$^{-1}$, while the outer part
of CSM is close to static.  During the propagation, the 
shock wave transfers its momentum to the CSM as seen in the 
profile at Day 3.
Because of the much smaller mass, CSM gets a velocity
of $1.5 \times 10^9$ cm s$^{-1}$ which is much higher than 
 $8 \times 10^8$ cm s$^{-1}$ in the He star. Most of the star has already
developed homologous expansion after Day 5, except a small
non-vanishing reverse shock near the inner boundary of CSM.

\textit{Temperature (T)}: The compact pre-collapse He star has 
a surface temperature as high as $\sim 10^7$ K. The CSM
is assumed to be isothermal at $10^4$ K. When the shock wave reaches
the inner boundary of the CSM at Day 1, the surface temperature has 
already cooled down to 
$< 10^5$ K, while the shock heated matter can be as hot as
$2 \times 10^5$ K. At Day 5, the hot temperature bump smears out due
to diffusion.
Around Day 20, both the star and CSM become isothermal. 

\textit{Free~electron~fraction ($X_e$)}: $X_e$ is an important factor 
for the opacity since a free electron has a small but constant 
cross-section, compared to other
bound-free and bound-bound transitions, which strongly depend on the 
population of particular ionization stages. 
At the beginning, the 
star is completely ionized, while the CSM is less ionized 
with a fraction of $\sim 40 \%$.
Once the shock wave reaches
the surface the heat allows the rapid ionization of the CSM.
At Day 5, when the reverse shock is propagating backward, 
the matter near the outer layer of the star has a low enough density and temperature for recombination. At Day 10, the recombination 
region is more extended. The shocked heated front remains fully ionized 
but the whole profile is bumpy, because of the shock-induced
density fluctuations. At Day 20, when the photosphere starts to 
recedes, CSM cools down and 
more matter has recombined. At Day 30, the whole ejecta gradually
recombines. 

\textit{Optical~depth $(\tau)$}: 
Initially, the star and CSM are opaque, 
having $\tau > 1$ except at the
very shallow layer just below the stellar surface. As the star
expands, $\tau$ decreases.  However, the shock
front makes the opacity sufficiently high that the photosphere
remains at the outermost part of the star. By Day 20,
the photosphere gradually retreats to $M(r) \sim 40 ~M_{\odot}$. At Day 30 the photosphere
receded much deeper into the star, showing that the ejecta
gradually becomes transparent.

\textit{Luminosity ($L(r)$)}: 
When the shock starts to interact with CSM, the luminosity profile
becomes extremely bumpy because of the fluctuations of 
the temperature, density, free electron fraction, and thus opacity
because of shock propagation.
At Days 3 and 5 the stellar surface becomes the most brightest part of the 
star. However, the trough in luminosity develops when the reverse
shock allows an early recombination of atoms.
Beyond Day 20,
the luminosity profile has reached a steady 
state that radiation develops steadily and reaches a uniform
luminosity at Day 30. 
The photosphere recedes by $\sim 1 ~M_{\odot}$
per 10 days.

\subsection{Radiation Hydrodynamical Results}

\begin{figure*}
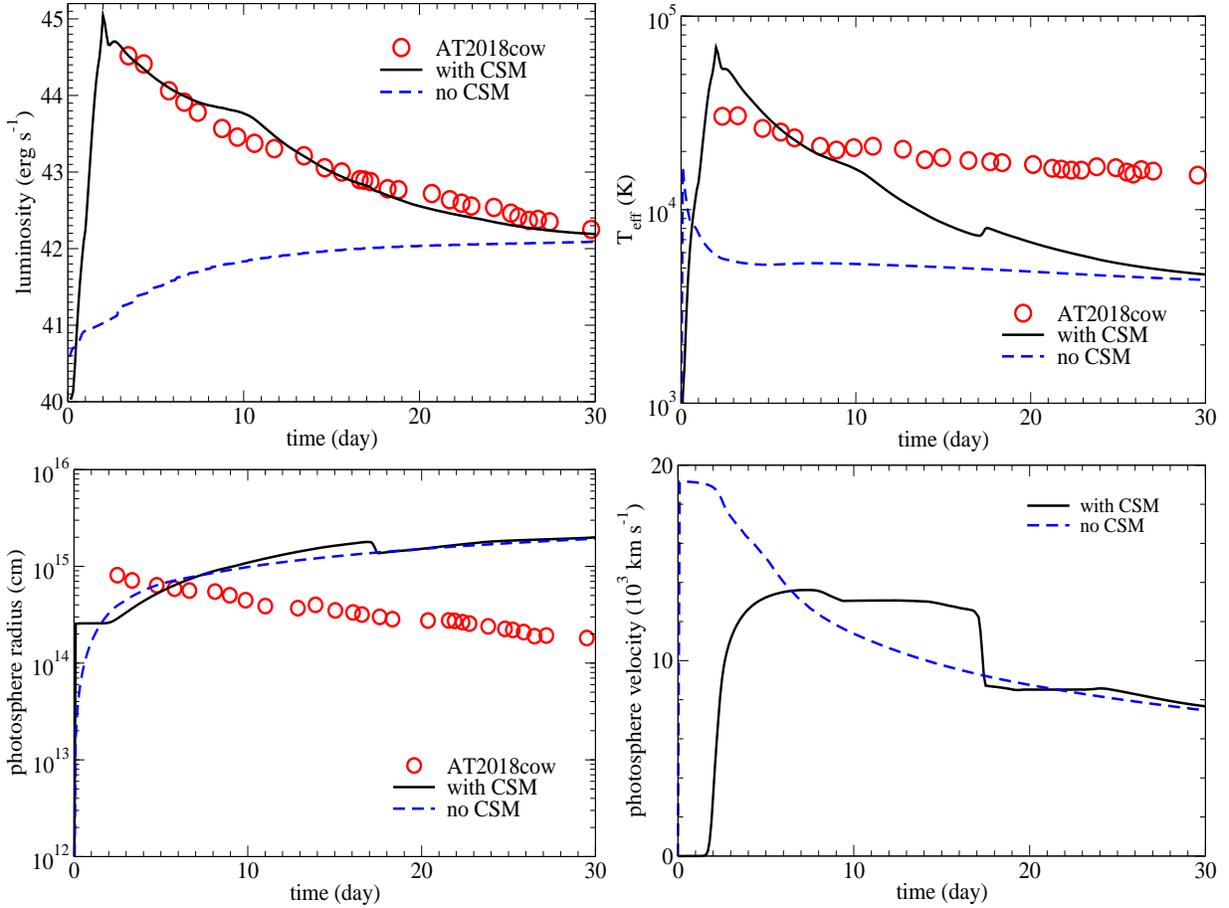

\centering
\includegraphics*[width=8cm,height=6cm]{lumin_time_plot.eps}
\includegraphics*[width=8cm,height=6cm]{Teff_time_plot.eps}
\includegraphics*[width=8cm,height=6cm]{photo_rad_time_plot.eps}
\includegraphics*[width=8cm,height=6cm]{photo_vel_time_plot.eps}
\caption{The luminosity (top left), effective temperature (top right),
photosphere radius (bottom left), velocity at the photosphere (bottom right) 
against time for the optimized model. The models with and without 
CSM predicted by the stellar evolutionary models are shown.
The data points correspond to those from AT2018cow \citep{Perley2018}.}
\label{fig:benchmark_rad_plot}
\end{figure*}

In Figure \ref{fig:benchmark_rad_plot} we plot the 
radiation hydrodynamical results of our ``optimized'' model, 
i.e., the bolometric luminosity (top left panel), 
effective temperature (top right panel), 
photospheric radius (bottom left panel) and 
velocity at the photosphere (bottom right panel).
The solid lines show the model with the circumstellar interaction as
described in the earlier subsection.
When the shock-breakout from the optically thick CSM occurs at Day 2,
the bolometric luminosity reaches the bright peak.  Then the
luminosity rapidly declines through radiative cooling.  Such a sharp rise
to a bright peak and a rapid decline of the light curve well
reproduce the observed FBOT-like feature of AT2018cow \citep{Perley2018}.
After day 18, the luminosity would decrease too rapidly, 
if there would be no radioactive heating as already shown in Figure
\ref{fig:benchmark_kappa_plot}.

For comparison, the results for the model without CSM
are shown by the dashed curve.
There is no sign of shock-breakout where the light curve is smooth and flat.
Its luminosity around Day 2 is
almost 3 orders of magnitude lower than the peak due to circumstellar interaction, 
but it reaches an asymptotic value of $\sim 10^{42}$ erg s$^{-1}$ at Day 20. 

Then the bolometric light curve produced by the combined powers of
circumstellar interaction and the radioactive decays shows a good
agreement with AT2018cow. The effect of shock heated CSM is seen in
the luminosity evolution.

However, the observed temperature and radius at the photosphere of
AT2018cow monotonically decrease \citep{Perley2018},
which suggests that the photosphere
recedes inward in $M(r)$ at a pace different from the model.  Such a
difference in the local quantities at the photosphere might stem from
a possible aspherical structure of CSM and the ejecta of AT2018cow.
It would be interesting to investigate the radiation hydrodynamics of
a multi-D structure of CSM-ejecta.

\subsection{Spectral Evolution of AT2018cow}

Although we focus on mainly reproducing the evolution of the
bolometric flux from AT2018cow,
our optimized model broadly outlines also some aspects of its spectral
evolution \citep{Prentice2018,Perley2018}.

According to the observations, the spectra of AT2018cow are very
blue and almost featureless
in the beginning, for days 4 to 8 after the maximum \citep{Perley2018},
which roughly corresponds to Days 7 to 11 after the explosion.
This is exactly what we expect to see in our optimized model.
The optical depth in Figure \ref{fig:benchmark_rad_plot} shows
that on Day 10 the photosphere is located in the outermost layers of
CSM.  These layers are shock-heated to about 20,000~K at this time.
The continuum emission must be very bright under such conditions,
and optical lines, even if some of them were formed at so high
temperature, sink under the continuum level.

Between Days 10 and 20, the photosphere starts to dive back into the
ejecta layers,
while the temperature falls down to 9,000--10,000~K.
The conditions become more suitable for forming optical lines,
and exactly during this period they appear in the spectrum of AT2018cow.

The problem arises with the explanation of weak narrow components of lines
that are often observed after Day 20 in the spectra of AT2018cow.
These lines do not appear in the ejecta and CSM of the optimized model,
because all materials in the model are already accelerated to velocities
as high as 7,000 - 13,000 km s$^{-1}$.
But these weak components do not resemble the typical shape of the
lines of Type IIn supernovae (SNe IIn).   
These are much weaker than in SNe IIn.
We suppose that these weak lines must not arise from the spherical envelope, into
which the exploding object is embedded,
as it happens in SN IIn's.  Instead, these lines 
can be emitted, for example, by more extended nearby 
(possibly a disk-shaped) structure
lost by the progenitor much earlier, having low velocity and
illuminated by the explosion.

The late appearance of hydrogen lines \citep{Perley2018} is expected 
by such extended CSM with hydrogen that was lost earlier (subsection
\ref{sec:csmFormation}).
As will be shown in
Section~\ref{sec:composition} and Figure~\ref{fig:rho_ini_tail_plot},
when we admix large amount of hydrogen into the helium CSM, this
affects the bolometric light curve very weakly.

\section{Dependence of Light Curve on Model Parameters}
\label{sec:sensitivity}

\begin{figure}
\centering
\includegraphics*[width=8cm,height=6cm]{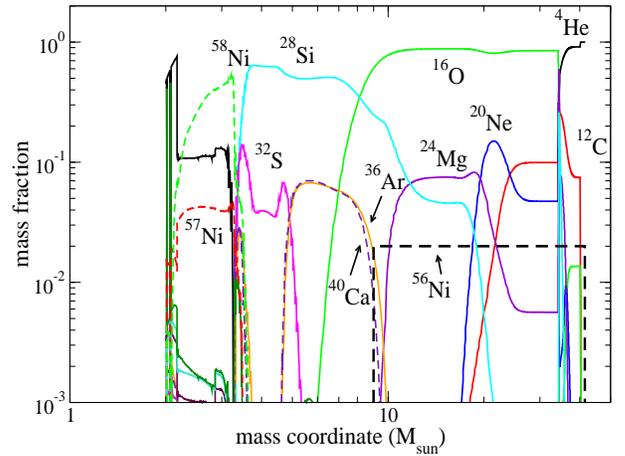}
\caption{The abundance distribution adopted in the ``comparison" model 
used for the parameter study in this section. It assumes 
that only radioactive $^{56}$Ni is brought to the outer layers of
$M(r) = 9.0 - 41.6 M_\odot$.}
\label{fig:partial_mix}
\end{figure}

\begin{figure}
\centering
\includegraphics*[width=8cm,height=6cm]{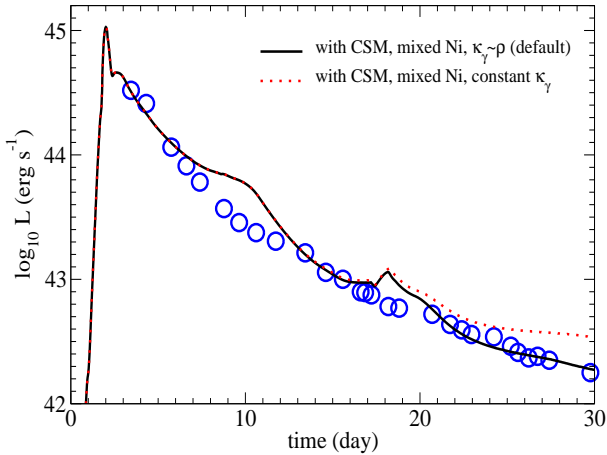}
\caption{The black-solid line shows the bolometric light curve 
  of the ``comparison" model based on the abundance distribution in
  Figure \ref{fig:partial_mix} (see text) as compared with the
  observed data of AT2018cow \citep{Perley2018}.  For radioactive
  heating, the black-solid line adopts $\kappa_{\gamma} = 0.06
  Y_{\rm e}$ g cm$^{-2} (\rho/\rho_0)$ cm$^2$ g$^{-1}$ if $\rho <
  \rho_0 = 10^{-10}$ g cm${-3}$.  If the constant $\gamma-$ray opacity
  $\kappa_{\gamma} = 0.06 Y_{\rm e}$ g cm$^{-2}$ is adopted, the light
  curve is shown by the red-dotted line, whose decline is too slow to
  be compatible with AT2018cow.  
}
\label{fig:benchmark_kappa_plot2}
\end{figure}

\subsection{The Comparison Model}
\label{sec:comparison}

In the previous section we have studied how the combined powers of
circumstellar interaction and radioactive decays 
can explain the bolometric light curves of AT2018cow
and obtained the ``optimized'' model.

We should note that the theoretical bolometric light curve depends on
a number of parameters and assumptions adopted in the modeling.  Here, we study how the light curves depend on the choice of these parameters.

For these comparison studies, we construct a ``comparison" model 
whose bolometric light curve is in fairly good agreement with
AT2018cow but with a different set of model parameters from the
``optimized'' model.  In this ``comparison" model, the elemental
abundance distribution in Figure \ref{fig:partial_mix} is assumed.
Here the original abundance distribution is adopted to calculate the
equation of state and opacities except for radioactive $^{56}$Ni which
is moved to the outer layers at $M(r) = 9.0 - 41.6 ~M_\odot$.  The aim
is to study the enhanced heating effects of $^{56}$Ni by mimicking the
jet-like ejection of $^{56}$Ni-rich region from a deeper layer.

In Figure \ref{fig:benchmark_kappa_plot2}, the bolometric light curve
for this ``comparison'' model is shown by the black-solid line
compared with the observed data of AT2018cow \citep{Perley2018}.  For
radioactive heating, the ``comparison" model (black-solid line)
adopts a $\kappa_{\gamma} = 0.06 Y_{\rm e}$ g cm$^{-2} (\rho/\rho_0)$
cm$^2$ g$^{-1}$ if $\rho < \rho_0 = 10^{-10}$ g cm${-3}$.  If the
constant $\gamma-$ray opacity $\kappa_{\gamma} = 0.06 Y_{\rm e}$ g
cm$^{-2}$ is adopted, the light curve is shown by the red-dotted line,
whose decline is too slow to be compatible with AT2018cow.
The reduction of $\kappa_{\gamma}$ at low densities could mimic
the effects of clumpy density distribution \citep[e.g., Eq.(1)
  of][]{Kumagai1989} in the ejecta.  Because the black-solid curve
is in better agreement with the late-time data of AT2018cow, we
treat this ``comparison" model as a default model for comparison.

\subsection{Dependence on Explosion energy}

\begin{figure}
\centering
\includegraphics*[width=8cm,height=6cm]{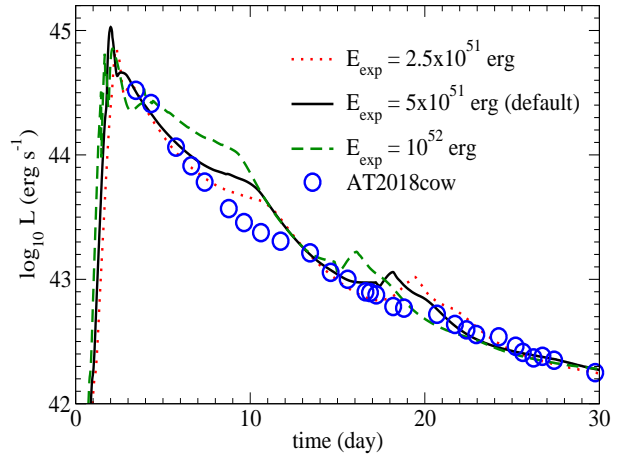}
\caption{Light curves of the comparison model with $E = 5 \times 10^{51}$ erg and its
variations with different explosion energy.
The data points correspond to those from AT2018cow. }
\label{fig:benchmark_Eexp_plot}
\end{figure}

In the light curve calculation, we treat the kinetic energy of
explosion $E$ as a model parameter.  Theoretically, the explosion
mechanisms of the core-collapse supernovae are not well-understand.
Thus it is not certain how much explosion energy is given to the
ejecta.  For example, if a black hole is formed with an accretion disk
around it, a powerful bipolar jet from magnetohydrodynamical
instabilities may provide a large amount of energy to the ejecta.
Without actual modeling of a core-collapse supernova, it would be
useful to constrain the explosion energy from the light curve
modeling.  Here we examine how the light curve depends on the explosion
energy.
We thus search the explosion energy from $E =$ $10^{51}$ to $10^{52}$
erg which produces the closest light curve to AT2018cow.

In Figure \ref{fig:benchmark_Eexp_plot} we plot the light curves for
the comparison model with an explosion energy of $E = 2.5 \times 10^{51}$,
$5 \times 10^{51}$ and $10^{52}$ erg, respectively. 
Other parameters, such as the CSM profile and the resolution are the
same as the comparison model.
It is seen that the effects of explosion energy is secondary. It does not play
an important role in the shock-breakout time and the peak luminosity. 
It affects a little the post-maximum decrease and the fluctuations of
the luminosity.  The model with a 
higher explosion energy fades slower, because the 
stronger shock makes the post-shock density higher, thus making
the recession of the photosphere slower.

\subsection{Dependence on CSM Structure}

\begin{figure}
\centering
\includegraphics*[width=8cm,height=6cm]{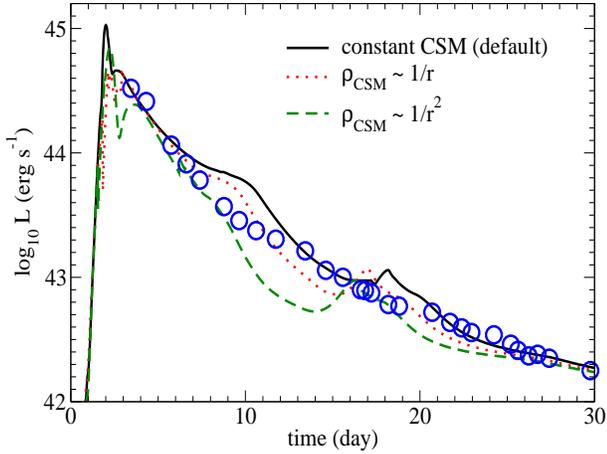}
\caption{Light curves of the default model and its
variations with different CSM density profiles.
The blue circles show the observed data of AT2018cow.}
\label{fig:benchmark_tail_plot}
\end{figure}

The comparison model is assumed to have 
CSM with a constant density of $10^{-11}$ g cm$^{-3}$.
Here we examine how the light curve depends on the CSM density
structure by adopting a profile of $(1 / r^{\alpha}$ with
$\alpha = $ 0, 1 and 2 as shown in Figure 
\ref{fig:rho_ini_tail_plot}.  In constructing the density profile, 
we require that CSM has the same mass $\sim ~0.5 ~M_{\odot}$.
and the same outer radius as the comparison model.

In Figure \ref{fig:benchmark_tail_plot} we plot the light curves for
the comparison model (black-solid) and its variations with different
CSM density profiles: i.e., $1/r$ (red dashed line) and $1/r^2$ (green
dot-dashed structure).
It is seen that qualitatively the light curve is not 
sensitive to the CSM density profile.  The steeper CSM density
results in a quicker decrease in the luminosity and 
a lower luminosity peak.
The late luminosity evolution becomes very close to each other
and overlaps when the $^{56}$Co decay dominates the light curve.

\subsection{Dependence on CSM Density}

\begin{figure}
\centering
\includegraphics*[width=8cm,height=6cm]{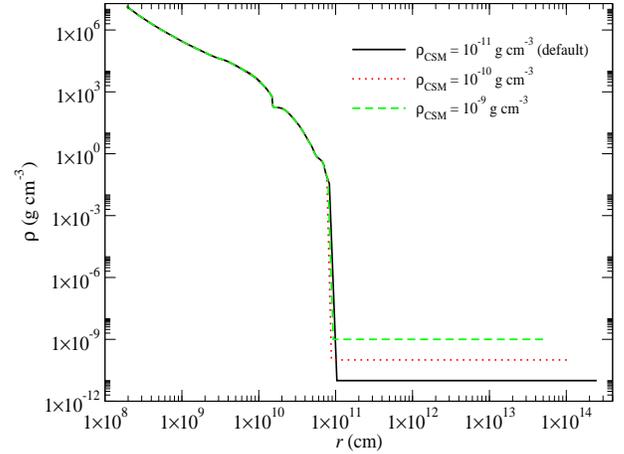}
\caption{The initial stellar and CSM density profiles of the models
comparing the effects of CSM density. All models have the 
same CSM mass 0.5 $M_{\odot}$ and a flat CSM profile.}
\label{fig:rho_ini_CSMrho_plot}
\end{figure}

\begin{figure}
\centering
\includegraphics*[width=8cm,height=6cm]{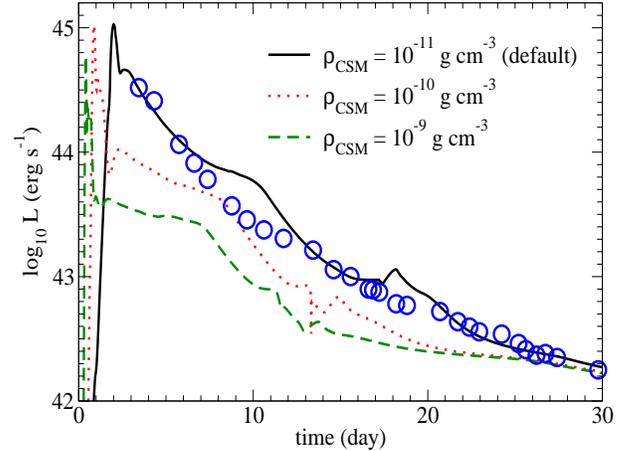}
\caption{Light curves of the default model and its
variations with different CSM density of $10^{-10}$ (red dotted line)
and $10^{-11}$ g cm$^{-3}$ (green dashed line).
The blue circles show the observed data of AT2018cow. }
\label{fig:benchmark_CSMrho_plot}
\end{figure}

Here we examine the dependence on the CSM density $\rho_{{\rm CSM}}$. 

We take the total mass of CSM is the same as comparison model
and vary the inner and outer boundaries of CSM as two model parameters.  
In Figure \ref{fig:rho_ini_CSMrho_plot} we plot the 
initial structure of the comparison model
and those with the various CSM density from $10^{-11}$ to $10^{-9}$ g cm$^{-3}$.
For a higher $\rho_{{\rm CSM}}$, the outer radius of CSM can be
almost an order of magnitude smaller than the model with a lower $\rho_{{\rm CSM}}$.

In Figure \ref{fig:benchmark_CSMrho_plot}, we plot the corresponding
light curves for the explosion models assuming the same 
explosion energy $5 \times 10^{51}$ erg and same resolution 
of mass grid 0.04 $M_{\odot}$ as the comparison model. 
The CSM density plays an important role in two aspects.

(1) First, the initial peak (shock breakout) depends sensitively 
on $\rho_{{\rm CSM}}$ and hence the outer radius of CSM. 
The first peak changes from Day 3 to Day 1 when 
$\rho_{{\rm CSM}}$ increases from $10^{-11}$ to $10^{-9}$ g cm$^{-3}$,
which corresponds to the decrease in the CSM outer radius from $3 \times 10^{14}$ cm 
to $5 \times 10^{13}$ cm.

(2) Second, the post-peak evolution
before the $^{56}$Co decay is sensitive
to $\rho_{{\rm CSM}}$. For lower $\rho_{{\rm CSM}}$,
the post-peak fall in the luminosity is faster. It takes about 5 days
for the comparison model ($\rho_{{\rm CSM}} = 10^{-11}$ g cm$^{-3}$)
to decrease one order of magnitude in luminosity, compared to the one with higher
$\rho_{{\rm CSM}} = 10^{-9}$ g cm$^{-3}$, for which it takes about 10 days 
for the luminosity to decrease by one order of magnitude.

\subsection{Dependence on Hydrogen in CSM}
\label{sec:composition}

\begin{figure}
\centering
\includegraphics*[width=8cm,height=6cm]{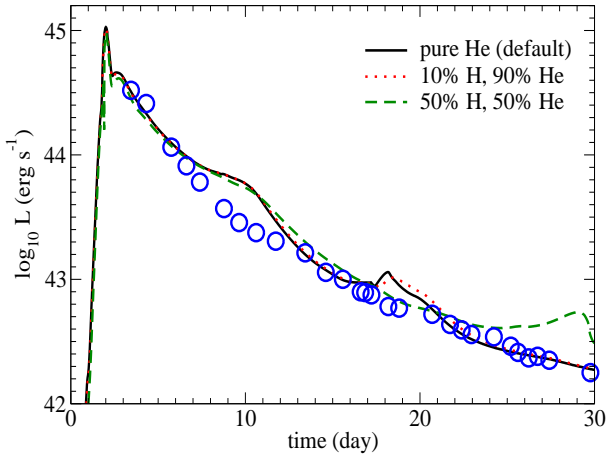}
\caption{Light curves of the comparison model and 
the models that includes some hydrogen in CSM (10 \% H and 50 \% H).   
The blue circles show the observed data of AT2018cow. }

\label{fig:benchmark_CSMcomp_plot}
\end{figure}

In the spectra of AT2018cow, H-features appear $\sim$ 15 days after
the light maximum \citep{Perley2018}.  
We then discuss in subasection \ref{sec:csmFormation} the formation
scenario to have some H in the ejecta and CSM.

To examine the possible effects of H in CSM, we approximate the
composition of CSM from pure He to two other abundances: a low H
mixing (0.1 H and 0.9 He) and a high H mixing (H and He are 50 \% by
mass fraction).

In Figure \ref{fig:benchmark_CSMcomp_plot} we plot the corresponding
light curve with the data point from COW. 
The light curves are almost identical 
despite the photosphere still lies inside CSM. 
The model with the higher H mixing has a slightly lower bolometric
luminosity at Day 18 despite a similar light curve shape.
The effects of H only become observable
in the H-rich model which shows a higher luminosity 
at Day 28 -- 30.

\subsection{Dependence on CSM Mass}

\begin{figure}
\centering
\includegraphics*[width=8cm,height=6cm]{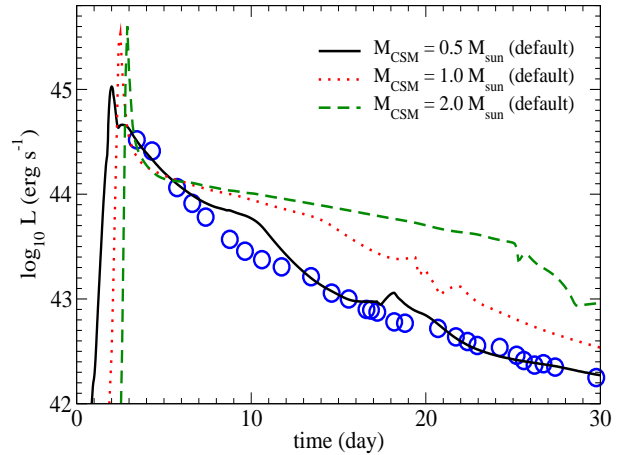}
\caption{Bolometric light curves for $M_{\rm CSM}/M_\odot =$ 0.50 
(black solid; comparison), 1.0 (red dotted), and 2.0 (green dashed).
$R_{\rm CSM}$ is fixed at $\sim 2 \times 10^{14}$ cm.
The blue circles show the observed data of AT2018cow. }
\label{fig:benchmark_MCSM_plot}
\end{figure}

The mass of CSM $M_{\rm CSM}$ depends on the 
dynamical mass loss during PPI, thus being sensitive to the progenitor mass.
Here we examine how the light curve depends
on $M_{\rm CSM}$ for the same $R_{\rm CSM}$ as of the comparison model.

In Figure \ref{fig:benchmark_MCSM_plot} we plot the light curves
for three models with $M_{\rm CSM}/M_\odot =$ 0.50 (black solid line for the comparison model),
1.0 (red dotted line), and 2.0 (green dashed line), 
having the same $R_{\rm CSM}$.

The light curve features depend strongly on $M_{\rm CSM}$. 
Larger $M_{\rm CSM}$ delays the shock breakout of the light curve
from Day 2 to about Day 3. The peak luminosity is also higher
for larger $M_{\rm CSM}$. The declining slope of the light 
curve becomes flatter when $M_{\rm CSM}$ is large. 
For $M_{\rm CSM} = 2.0 M_\odot$, the light curve
remains $\sim 10^{44}$ erg s$^{-1}$ between day 5 -- 25. 
As a result, the transition to the phase powered by $^{56}$Co-decay 
is also slightly delayed. These trends are the effects of 
a longer diffusion time in more massive CSM.

Our results confirm that $M_{\rm CSM}$ is strongly constrained by the
light curve width, and as small as $\sim 0.5 M_\odot$ to be consistent
with AT2018cow as \cite{Perley2018} has estimated.

\subsection{Dependence on Resolution}

\begin{figure}
\centering
\includegraphics*[width=8cm,height=6cm]{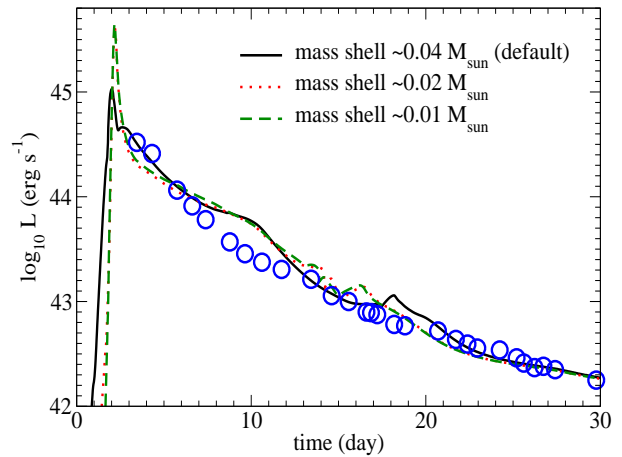}
\caption{Light curves of the comparison model and its
  variations with different resolutions of the mass
  coordinate.}
\label{fig:benchmarkresolution_plot}
\end{figure}

How the shock propagates through the CSM controls the 
early evolution of the light curve. That means how the
shock is resolved is important to trace how the 
kinetic energy of the shock is transformed to the 
internal energy and hence the blackbody radiation
through shock compression. 
Here we examine how the results depend on the choice of the 
resolution. 

In Figure \ref{fig:benchmarkresolution_plot} we plot
the light curves of the comparison model with 
the current resolution (1000 grids), two times
and four times higher resolutions.  The highest resolution 
has a mass resolution of $\sim 10^{-2} ~M_{\odot}$.
The qualitative features 
of the light curve are very well captured from
the current resolution. The higher resolution
has a higher maximum luminosity in the early peak, owing to the 
smaller mass shell. After the shock breakout, 
three models behave similarly, except when 
Day 20 the bump appears later
when the resolution is lower. Despite that the similarity
of the three light curves shows that the shock propagation
is already well captured by the current resolution.

\subsection{Comparison with {\sc STELLA}}
\label{sec:STELLA}

In this subsection we compare the numerical results with the
multi-color radiative transfer calculation done by the code {\sc stella}
\citep{Blinnikov2006,Baklanov2015}.
In the bulk of the present work we did not use {\sc stella} 
because its
implicit and multi-band nature makes the parameter searching
for the optimized light curve unavoidably time-consuming.

Here we compare our results obtained from the 
blackbody diffusion-limit radiative transport calculation by {\sc snec}
with the results computed by {\sc stella}.
We give comparisons only for the bolometric light curve of the comparison model
described in subsection \ref{sec:comparison}.
The multi-band light curves calculated by {\sc stella} for another
set of models are presented
in a separate publication (Sorokina et al., in preparation).

\begin{figure}
\centering
\includegraphics*[width=8cm,height=6cm]{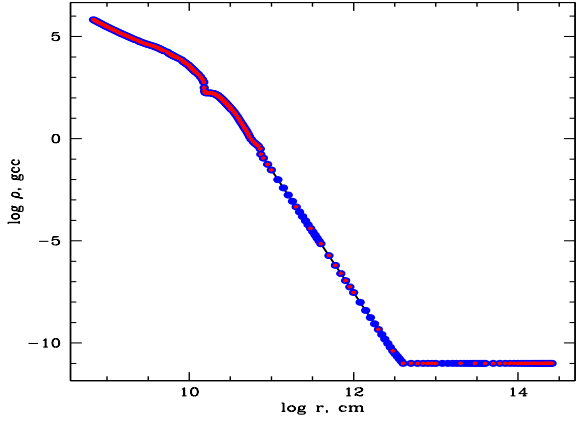}
\caption{Initial density profile of the presupernova model for comparison (red
  stars) and its remapping for {\sc stella} runs (blue dots). }
\label{fig:preSN_STELLA}
\end{figure}

The presupernova model calculated by MESA has more than 1200 mesh
zones.  Blue dots in Figure~\ref{fig:preSN_STELLA} show our remapping
of this model for {\sc stella} runs with 232 zones, which is much more
uniform than the one used for {\sc snec}.  It is clear that the
resolution of the structure in interior is sufficiently high.

The both codes {\sc stella} and {\sc snec} are spherically symmetric
Lagrangean radiation-hydrodynamic ones.  Hydrodynamics equations
embedded in {\sc stella} and {\sc snec} codes are quite similar.  The
principal differences between the codes are in the implementation of
radiative transfer into hydrodynamical simulations.
{\sc stella} solves implicitly time-dependent equations for the angular
moments of intensity in fixed frequency bins  which are coupled with
the Lagrangian hydrodynamical equations.
Therefore, there is no need to ascribe any temperature to the
radiation, thus the photon energy distribution may be quite arbitrary.
{\sc snec} uses the equilibrium-diffusion approximation for radiation
transport. Thus a blackbody spectrum is enforced in {\sc snec} with
the same temperature for radiation and matter.

\begin{figure}
\centering
\includegraphics*[width=8cm,height=6cm]{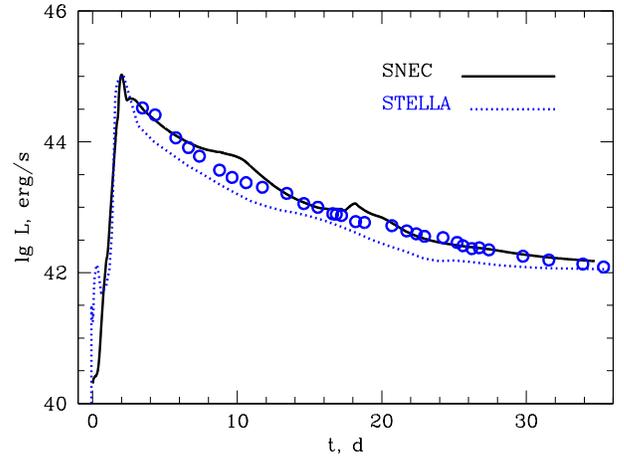}
\caption{Bolometric light curves adopted by {\sc snec} (thin line) and {\sc stella} (dotted
blue line) for the presupernova models shown in
Figure~\ref{fig:partial_mix}.
}
\label{fig:SNECvsSTELLA}
\end{figure}

In the {\sc snec} model, the photosphere is located in the outer layers 
during the first 35 days, and the black-body approximation is
quite applicable to the reconstruction of the bolometric light curve.
Figure~\ref{fig:SNECvsSTELLA} demonstrates a reasonably good agreement
of the bolometric fluxes of {\sc snec} and {\sc stella} for 
this period.

We observe that the two codes agree qualitatively well.  The two
models can produce the rapid rise of AT2018cow within the
first 3 -- 4 days.  Later the luminosity in {\sc stella} falls a bit
faster due to a faster recombination in the outer layers with the mass
$\sim 2 ~M_\odot$ from the edge of the ejecta which are above the
photosphere at this epoch.  Given many differences in the treatment of
opacity and radiative transfer in {\sc snec} and {\sc stella} we find
that the agreement is satisfactory.  Understanding the effect on light
curves of different approaches in {\sc snec} and {\sc stella} codes
requires a detailed comparative study (Sorokina et al., in
preparation).

Note {\sc stella} calculates both the effective temperature $T_{\rm
  eff}$ and the color temperature $T_{\rm color}$.  At the shock
breakout, $T_{\rm color}$ reaches $\sim 10^6$ K, which is much higher
than $T_{\rm eff} \sim 10^5$ K obtained by {\sc snec}.

\section{Discussion}
\label{sec:discussion}

In the previous sections, we have shown that the FBOT-like feature of
the early bolometric light curve of AT2018cow is well reproduced by
circumstellar interaction in our optimized model before Day 20.  After
Day 20, the bolometric light curve is well-modeled by radioactive
decays, but it requires extensive mixing of $^{56}$Ni almost uniformly
up to the ejecta-CSM interface.
Thus it is still worth discussing other possible central energy
sources to power the late light curve in our PPISN model.

\subsection{Fallback Onto Black Hole}

\begin{figure}
\centering
\includegraphics*[width=8cm,height=6cm]{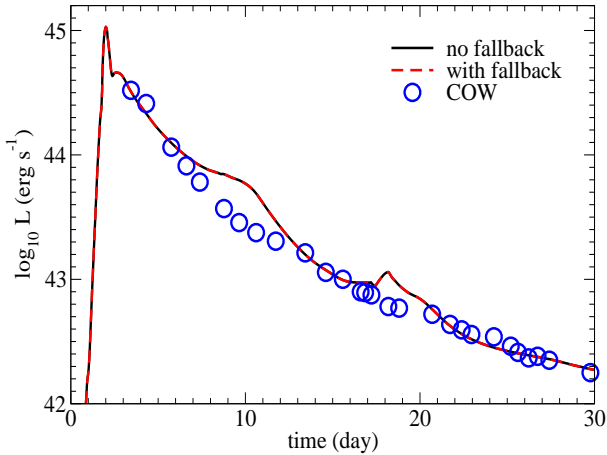}
\caption{Light curve of the comparison model and 
variations with different CSM composition without or with
fallback energy source
The data point corresponds to AT2018cow. }
\label{fig:benchmark_TDE_plot}
\end{figure}

The progenitor of PPISN is as massive as 80--140 $M_\odot$, thus
it is likely that a black hole is formed in the center.
It is possible that fallback 
of matter onto the black hole occurs to provide a late phase energy
source other than the $^{56}$Ni decay
or circumstellar interaction.
In view of that, we include the fallback energy source in the comparison model.
In our spherical explosion model,         
where we insert an energetic thermal bomb,
no direct fallback occurs, so we use
a parameterized fallback formula. 

We adopt the analytic formula for energy deposition \citep{Michel1988,Chevalier1989,Dexter2013}
\begin{equation}
\dot{\epsilon} = A / M_{{\rm dep}} t^{-2.5}, 
\end{equation}
where $A$ is a parameter for the energy deposition rate, and
$M_{{\rm dep}}$ is estimated by the photon mean free path. However the
compact inner core most of the time leads to energy deposition focused
in innermost shells for most of the time. $A$ is determined by how
much mass is accreted during the simulation. The new model has two
energy sources: one is the decay of $^{56}$Ni and the other is the
fallback.



In Figure \ref{fig:benchmark_TDE_plot} we plot the light curves for the 
two models, and find that the two light curves overlap with each other. 
This is expected because, in the early light curve, the photosphere
is located in CSM or in the outer ejecta. The energy from fallback
has indeed modified the internal structure of the core by
thermal expansion. However, the timescale for the energy 
to arrive at the photosphere depends on the diffusion timescale.
As a result, no significant change can be seen 
in the model with the fallback energy deposition.

However, we remark that in multi-dimensional modeling of 
fallback can be very different from the one-dimensional model.
The jet structure may form when the progenitor rotates. 
The rotating black hole accommodates a rapidly rotating
accretion disk, where the gravitational instability in the 
disk after the formation of dead zone can trigger large-scale
mass and energy ejection \citep[see e.g.][]{Tsuruta2019}. 
In a multi-dimensional model \citep[see, e.g.,][]{Tominaga2007}, 
the jet can break out and form two holes. 
This drastically reduces the 
the site of the fallback onto the black hole
to the photosphere. To reproduce this phenomenon, we need
to parameterize the energy deposition depth into the outer ejecta.

\subsection{Magnetar Model}

We do not consider a magnetar \citep{Kasen2010,Kasen2016}
as a power source for the late light
curve in the present PPISN model because of the following reasons.

(1) The progenitor of PPISN is so massive as 80 -- 140 $M_\odot$ that it
may not form a neutron star remnant.
(2) Even if we assume the formation of a magnetar in our PPISN model,
the diffusion time for the magnetar energy to reach the photosphere
would be too long to explain the observed light curve after Day 18,
being similar to the black hole accretion model as shown in Figure
\ref{fig:benchmark_TDE_plot}.

Note, however, the magnetar activity may form a jet-like ejecta,
The magnetar, which loses energy through its dipole radiation, can
effectively transfer its energy by electromagnatic waves along the
confined angle.  Such a jet-like structure 
might reduce the timescale to transfer the magnetar energy
to the surface.

If the progenitor of AT2018cow would be much smaller star, however, it
may not encounter the above two problems \citep[e.g.,][]{Fang2019}.
In such a low mass model, we need a circumstellar matter to power the
early light curve to reproduce the FBOT-like feature.  Actually, the
super-AGB progenitor of the electron capture supernova forms such CSM
as well as a neutron star, and that is the model applied for FBOT, KSN
2015K, by \cite{Tolstov2019}.  Such a case would be worth
investigating for a model of AT2018cow (Sorokina et al. in
preparation).

\subsection{High Energy Photons of AT2018cow}
\label{sec:highenergy}


In \cite{Margutti2019} the detailed X-ray and gamma-ray
light curve and spectra of AT2018cow are presented.

Here we discuss how the possible energy source of the light curve of
AT2018cow (circumstellar interaction, magnetar and accreting BH) can
be related to the detected X-ray.

In the circumstellar interaction model, the shock-heated matter has
the ``color" temperature as high as $\sim 10^6$ K according to the
calculation of STELLA as mentioned in subsection \ref{sec:STELLA}.
Such high temperature materials emit X-rays, which may easily escape
from the star.

If the aspherical explosion is the case as illustrated in Figure 12 of
\cite{Margutti2019}, the ejecta near the "equator" can be ejected
faster and is less compact than those near the "poles".  The formation
of aspherical circumstellar material allows time-lapse for the
interaction to take place.

When bipolar-like explosion takes place, the two poles are more
readily to be penetrated by the high velocity flow. Depending on the
jet energetics, the jet can breakout directly the surface and trigger
the X-ray burst by the interaction.

If the aspherical explosion occurs, the 
bipolar structure forms as discussed in earlier subsection for
the accreting black hole and magnetar, and allows early X-ray emission.

When the shock breakout occurs, it creates an opening 
of the star reaching directly the compact core. The high-energy
photon coming from the black hole or neutron star can escape the 
star efficiently. 

The exact nature of the aspherical explosion and its early high-energy
photon emission (see Figure 12 of \cite{Margutti2019}) will require
multi-dimensional radiative transfer simulations for understanding,
which will be an interesting future project.

\section{Conclusion}
\label{sec:conclusion}

In this work we apply the model of circumstellar interaction in the
pulsational pair-instability supernova (PPISN) for explaining the Fast
Blue Optical Transient (FBOT) AT2018cow (COW).  AT2018cow has quite a
unique early light curve, showing a peak as bright as SLSNe but much
faster rise and fall than SLSNe.  We apply the evolutionary model of
the $42 ~M_{\odot}$ He star which collapses after undergoing PPI mass
ejection, and compute their corresponding bolometric light curves. We
have searched for the optimized model parameters (explosion energy,
CSM density and structure, and distribution of $^{56}$Ni) with which
the bolometric light curve of AT2018cow is well-reproduced.

We show that an explosion of PPISN with the energy $\sim 5 \times
10^{51}$ erg, a $^{56}$Ni mass of $\sim 0.6 ~M_{\odot}$, CSM mass of
$0.5 ~M_{\odot}$ and a density of $10^{-11}$ g cm$^{-3}$ produce an
optimized model whose bolometric light curve is in best agreement with
AT2018cow. 

We also studied how each model parameter affects the light curve.
We note that the simulation reaches the convergence regime in the
resolution (mass shell $\sim 2 \times 10^{31}$ g).  The explosion
energy plays a secondary role in the light curve shape. On the other
hand, the CSM mass, density, and the structure dominate the light
curve shape. Mixing of $^{56}$Ni is necessary to explain the slow
decline of the luminosity beyond Day 20.  Observable differences in
the photosphere evolution suggests that further validations are
necessary to connect the FBOT AT2018cow to PPISN model.

To explain the late time light curve and the high-energy photon after
shock breakout, a central engine such as fallback onto a black hole or
a magnetar remains important. Despite that, the interaction can still
provide the necessary condition for the rapid rise and drop of the
light curve up to $\sim 40$ days. Further discrimination of models
will require multi-dimensional simulations to trace how the aspherical
energy deposition contributes to the high energy photons detected.

Based on our successful model for FBOT (AT2018cow) with the CSM mass
of 0.5 $M_{\odot}$ and the model for SLSN \citep{Tolstov2017} with the
CSM mass of $(\sim 20 M_{\odot})$, we set the following working
hypothesis that PPISN produces SLSNe if CSM is massive enough and
FBOTs if CSM is less than $\sim 1 M_{\odot}$.

\section{Acknowledgment}

This work was supported by World Premier
International Research Center Initiative
(WPI Initiative), MEXT, Japan, and JSPS KAKENHI Grant Number JP17K05382 and JP20K04024.
K.N. would like to thank Brian Metzger and Raffaella Margutti for
useful discussion at the UCSB/KITP workshop "The New Era of
Gravitational-Wave Physics and Astrophysics" in 2019.
S.C.L thank the MESA development
community for making the code open-sourced and
V. Morozova and her collaborators in
providing the SNEC code open source.
S.C.L. acknowledges support by funding HST-AR-15021.001-A and 80NSSC18K1017.
S.B. is sponsored by grant RSF 19-12-00229 in his work 
on the supernova simulations with STELLA code.
P.B.'s work on understanding the effect on light 
curves of different approaches in SNEC and STELLA 
codes is supported by the grant RSF 18-12-00522.
E.S. is supported by the grant RFBR 19-52-50014  
in her work on developing codes modeling the radiative transfer in SNe.

\bibliographystyle{aasjournal}
\pagestyle{plain}
\bibliography{biblio}


\end{document}